\documentclass[12pt]{article}

\newcommand{\be}{\begin{equation}}
\newcommand{\ee}{\end{equation}}
\newcommand{\br}{\begin{eqnarray}}
\newcommand{\er}{\end{eqnarray}}

\def\beq{\begin{equation}}
\def\eeq{\end{equation}}
\def\bnq{\begin{eqnarray}}
\def\enq{\end{eqnarray}}
\def\barr{\begin{array}}
\def\earr{\end{array}}

\def\be{\begin{equation}}
\def\ee{\end{equation}}
\def\ba{\begin{eqnarray}}
\def\ea{\end{eqnarray}}

\def\br{\begin{array}}
\def\er{\end{array}}
\def\bc{\begin{center}}
\def\ec{\end{center}}

\def\lapp{\mathrel{\rlap{\raise.5ex\hbox{$<$}}
                    {\lower.5ex\hbox{$\sim$}}}}
\def\gapp{\mathrel{\rlap{\raise.5ex\hbox{$>$}}
                    {\lower.5ex\hbox{$\sim$}}}}
\def\DESepsf(#1 width #2){\epsfxsize=#2 \epsfbox{#1}}
\usepackage{color}
\begin{document}
\thispagestyle{empty}
\vskip 10pt 
\begin{center}
{\Large\bf Radiative seesaw in $SO(10)$ with dark matter}

\vskip .25in

{ {Mina K. Parida}}\\
{\bf{Harish-Chandra Research Institute, Chhatnag Road, Jhusi, 
Allahabad 211019, India.}\\
{Email: paridam@mri.ernet.in}}\\
\end{center}

\begin{abstract} High energy accelerators
may probe into
the dark matter and the seesaw neutrino mass scales if they are not much 
 heavier than  $\sim $O (TeV).  
 In the absence of supersymmetry, we extend a class of minimal $SO(10)$ models to predict  
well known cold dark matter 
candidates while  achieving precision 
 unification with experimentally testable proton lifetime. The most important 
 prediction is a new radiative
seesaw formula of Ma type accessible to
 accelerator tests while the essential small value of its quartic coupling  also emerges naturally. 
This dominates over the 
 high-scale seesaw
 contributions making a major
   impact on neutrino physics and dark matter,  
 opening up  high prospects as a theory of fermion masses.  
\end{abstract}

\noindent{\large\bf Introduction.} Over the recent years, there has been a
continued surge of interests in exploring the origin of dark matter of the universe 
while global efforts for understanding very small masses and large mixings in
the neutrino sector have been intensified. The discovery of
dark matter (DM) dates back to 1933 when, from velocity
measurements in the Coma cluster, Zwicky predicted the inevitable presence of large
clumps of massive nonluminous matter \cite{zwicky} which has been reconfirmed
by a number of astrophysical and cosmological observations including the 
WMAP \cite{wmap}. Based upon the gauge group 
$SU(2)_L\times U(1)_Y\times SU(3)_C (\equiv G_{213}) $ the standard model (SM)
predicts all neutrinos to be massless and no DM candidate.
More than 70
years ago, Majorana conjectured neutrinos to be their own anti-particles and
a neutrino mass may signify its Majorana character uncovering the violation of  well known symmetry called
the lepton number \cite{majorana}.
 The revelation of tiny
neutrino masses, intimately related to the neutrino oscillation phenomena which
was at first hinted through Davis' Cl-37 experiment \cite{davis} in 1964, has been ultimately confirmed by
atmospheric, solar, and reactor neutrino experiments \cite{rnmsm}. 
Nearly four decades ago non-supersymmetric (non-SUSY) grand
unified theories (GUTs) were proposed to unify three basic forces of nature 
with neat and robust prediction 
for proton decay, $p\to e^+\pi^0$, for which there are ongoing search
experiments \cite{ps,nishino}.
 Out of all GUTs,  $SO(10)$ has grown in
popularity as it can predict the right order of tiny neutrino masses through a 
path breaking new mechanism,
called the canonical ($\equiv$ type-I) seesaw mechanism, shown to be possible
only if neutrinos are Majorana fermions \cite{type-I}. To mention a few out of
a number of other qualities, while the model 
can explain the
origins of parity (P) and CP  violations, it has the potential for fitting all
fermion masses and also explain baryon asymmetry of the universe via
lepton asymmetry and sphaleron effects \cite{lepto}. While the heavy 
right-handed (RH) neutrinos in $SO(10)$ mediate the canonical seesaw, the same
theory also predicts
another seesaw formula  ($\equiv$ type-II) \cite {type-II}
but now mediated by a massive left-handed (LH) Higgs scalar triplet,
${\Delta}_L$, and the two mass formulas are
\ba
  {\rm M}_{\nu}^{I} = ~-{\rm M}_{D}{\rm M}_{R}^{-1}{\rm M}_{D}^{T},
~~{\rm M}_{\nu}^{II}&=&fv_L, \label{seesaw12}
\ea
where  
$v_L=\lambda {\rm V}_{R} v^2/{\rm M}_{{\Delta}_L}^2$, $v=$
standard Higgs vaccum expectation value (VEV),
 ${\rm V}_R =$ VEV of RH Higgs triplet $\Delta_R$, ${\rm M}_R=f{\rm V}_R
=$ right-handed (RH) neutrino
mass, ${\rm M}_D=$ Dirac mass of 
neutrino , and  ${\rm M}_{{\Delta}_L}=$  mass
of ${\Delta}_L$.

Available data on light neutrino masses constrain these scales to be high, 
$10^{13}-10^{15}$ GeV 
and, as such, large hadron collider (LHC) and future high energy accelerators
can not test the underlying origin of neutrino masses. Further, the minimal
non-SUSY $SO(10)$ fails to fulfil the very purpose of unifying the SM gauge
couplings for which it was designed, nor can it explain
the dark matter phenomena.\footnote{ The well known minimal $SO(10)$ model
is defined to 
be the one with standard fermion representation and Higgs representations
necessary to implement the desert type spontaneous breaking and seesaw
mechanisms.} 
However, supersymmetric (SUSY) $SO(10)$ with inbuilt Fermi-Bose symmetry
achieves almost precision
unification \cite{susy1} and predicts dark matter with potential for TeV scale
seesaw mechanism \cite{others}. 
 But SUSY GUTs have their own shortcomings too \cite{comp}. In any case, in the
absence of any  evidence of SUSY so far, it is worthwhile to
explore prospects of non-SUSY SO(10) while preserving precision gauge coupling
unification and dark matter as the twin guiding principles. 

In this Letter we show how the minimal non-SUSY $SO(10)$ model is extended to predict DM and
achieve precision unification with testable proton stability. With
matter parity conservation, while type-I and type-II seesaw are automatic
consequences of the model, it
also generates a low-scale radiative seesaw formula of Ma type \cite{rad3} accessible to accelerator
tests and this formula dominates over the conventional ones
causing a  major impact on neutrino physics and dark matter phenomenology 
 opening up high prospects as 
 a theory of fermion masses in general. Comparison of prototon lifetime
 predictions of the present $SO(10)$ model and the $S(5)\times Z_2$ model
 \cite{rad4} with the current experimental limit \cite{nishino} for $p\to e^+\pi^0$ reveals clear
 distinction between the two models. 

\noindent{\large\bf Precision unification.}~Prospective DM candidates  are
usually accommodated in model extensions by imposing additional discrete
symmetries for their stability. But
an encouraging aspect of non-SUSY $SO(10)$ is that 
when its gauged ${\rm U}(1)_{\rm B-L}$ subgroup  
breaks spontaneously by the same mechanism as the canonical seeesaw through the
high scale VEV of the right-handed Higgs triplet, $\Delta_R$, carrying $({\rm B-L})=-2$, the surviving matter
parity, ${\rm P}_{\rm M}=(-1)^{3({\rm B-L})}$, emerging as gauged discreet
symmetry  $Z_2$  \cite{gds}
  can safeguard the stability of DM candiadtes once the latter are
introduced into the model Lagrangian. The $SO(10)$
representations  $10, 45, 54,
120, 126$, and $210$ possess even matter parity, but the representations $16, 144, ...$ have odd matter parity, irrespective of whether they represent
fermions or scalars. Consistently, the SM 
fermions (Higgs) carry odd (even) matter parity. Therefore, the general 
principle for prospective DM particles is that, subject to fulfilment of all other
phenomenological constraints, they might be nonstandard fermions of even
${\rm P}_{\rm M}$ or scalars of odd ${\rm P}_{\rm M}$. Using suitable
extensions of minimal non-SUSY $SO(10)$
while the hyperchargeless weak triplet fermion with
well investigated phenomenology \cite{cirelli} has been predicted 
\cite{frigerio,psb}, independently, the inert scalar doublet has
also emerged as a CDM candidate \cite{kx}. But as we find
here, both the inert doublet and the fermion triplet can be made light, in
addition to other non-standard fermions of 
 $SO(10)$ leading to a substantial impact on neutrino physics, 
DM phenomenology,
 proton decay, and fermion masses.

For precision unification, at first we take out the scalar superpartners
of quarks and leptons from the well known spectrum of the minimal
supersymmetric standard model (MSSM). Then the remaining non-standard degrees of freedom due 
to the Higgs ($\chi$) and fermions ($F_i$) in the non-SUSY model at 
low scale are \footnote{ This may be recognized as the well known low scale
  spectrum in the split-SUSY model \cite{arkani} except for the presence of an
  additional Higgs doublet.}
\ba
\chi(2,1/2,1), F_{\phi}(2,1/2,1),F_{\chi}(2,-1/2,1),
F_{\sigma}(3,0,1), F_b(1,0,1),F_C(1,0,8).\label{add}
\ea
In eq.(\ref{add}) $F_{\phi}, F_{\chi}$ are
analogues of two Higgsino doublets,  $F_{\sigma}, F_b$ and $F_C$ are the
analogues of wino, bino and gluino, and all the fermions  
 except the octet  have been treated as potential
 CDM candidates in SUSY GUTs.
It is well known that without scalar superpartners, the fields in
eq.(\ref{add}) maitain unification of gauge coupling almost at the same scale
and  with the same level of
precision as the MSSM but with a decreased value of the  unification coupling. 
 We note that
when any one of the fields in eq.(\ref{add}) with nontrivial
quantum numbers is treated to be absent or made superheavy, the accuracy of 
precision unification at that scale is more or less reduced as in  
\cite{kx} while different combinations of fermions and 
scalars, but with exactly equivalent degrees of freedom as in eq.(\ref{add}),
 yield 
unification with the same precision as in the MSSM as shown in ref. \cite{psb}\footnote{Unification of
  couplings with radiative seesaw and triplet DM has been also discussed
  outside $SO(10)$ with assumed discrete symmetry \cite{rad4}.}.

Using the SM particle masses and 
 $m_{F_{\phi}}\simeq m_{F_{\chi}}= 2$ TeV,
 $m_{F_{\sigma}}\simeq m_{\chi} \simeq 3$ TeV, $m_{F_C}\simeq 6$ TeV,
the resulting precision unification of gauge couplings in the non-SUSY theory occurs 
 close to the MSSM GUT scale with  ${\rm M}_U =~10^{15.96}$ GeV , $\alpha_G^{-1} =
 35.3$. The closeness of the three couplings at the GUT scale is impressive,
with $\alpha_1^{-1}(M_U)= 35.34, \alpha_2^{-1}(M_U)=35.32$, and $\alpha_3^{-1}(M_U)=35.30$
where $\alpha_i=g_i^2/(4\pi)$. 
The precision unification is guaranted 
 in the presence of
SM gauge symmetry below the GUT scale by assuming the superheavy Higgs 
components in
each SO(10) representation to be degenerate in masses not very different from
the GUT scale which reduce the GUT-threshold effects considerably \cite{ds, parpat}. 

Like the SM fields, if the light fields given in eq.(\ref{add}) can also be shown to emerge 
 from suitable 
$SO(10)$ representations, then the non-SUSY GUT would be said to have 
realized the low-scale spectrum and this precision unification. 
For this purpose 
we  exploit the non-standard fermionic 
 representations, $ {45}_F(+)$ and ${10}_F(+)$, in addition
to the  Higgs representations $ {10}_H(+),{\overline {16}_H}(-),{45}_H(+), 
{\overline {126}_H}(+), {54}_H(+)$, and ${210}_H(+)$ where the respective matter 
parity, ($+$) or ($-$), has been  shown against each representation.
While the three fermions $F_{\sigma}, F_b, F_C\subset {45}_F$,  the
fermion-doublet pair,
 $F_{\phi}, F_{\chi}\subset {10}_F$. 
In order to make these nonstandard fermions 
light, we utilize the 
GUT-scale Yukawa Lagrangian
\ba
-L_{\rm Yuk}&=& {45}_F\left(m _{{45}_F} + \lambda_P{210}_H +\lambda_E
{54}_H\right){45}_F
+{10}_F\left(m_{{10}_F}+\lambda_P^{\prime}{45}_H+\lambda_E^{\prime}{54}_H\right){10}_F,
\label{gyuk}
\ea
where generation indices have been suppressed.
Utilizing the SM singlet VEVs in ${54}_H$, ${45}_H$, and ${210}_H$, we 
find that
the model has enough parameter space to make the
 four non-singlet fermions of eq.(\ref{add}) and two singlet fermions 
$F_b^i ~(i=1,2)$ of the second and the third generations light by suitable tuning of 
 the parameters in eq.(\ref{gyuk}) \cite{fukuyama} while safeguarding precision
 unification in the same fashion as shown in ref.\cite{psb}. We will show that $F_{\sigma}$ and $F_b^i~(i=1,2)$ effectively replace
 the roles of RH neutrinos in driving the radiative seesaw. 
Similarly the non-standard 
inert doublet $\chi(2, 1/2,1) \subset {\overline {16}_H}$
 is brought to the $\sim$O(TeV) scale \cite{dpar}. The presence of light fermions at
low scales may be natural in non-SUSY GUTs as their masses could be
protected by corresponding global symmetries. Perturbative and non-perturbative
resolutions of cosmological relic density problem that might otherwise 
arise due to TeV scale mass of color octet fermion
have been discussed earlier \cite{psb, baer}.   

To examine the impact of conserved matter parity on neutrino mass formulas 
we note that while the canonical and the type-II seesaw are automatic consequeces of this
model, a number of other types of  
formulas  normally  allowed in
the SM extensions or different SO(10) models
\cite{others} are now disallowed since, in  the following Yukawa interaction,
\be
-L^{\prime}_{\rm Yuk}=Y{16}_F{45}_F{\overline {16}_H}, \label{t3yuk}
\ee
 the matter-parity violating VEV of
 ${\overline {16}_H}$ is forbidden.
 Further, matter-parity
 conserving type-I and type-II 
 seesaw continue to remain as the only two formulas if,
 in eq.(\ref{add}), the second Higgs doublet $\chi \subset {10}_H(+)$ and
 carries even matter parity.   
 
\noindent{\large\bf Radiative seesaw.} ~The complexion of neutrino mass 
changes drastically once the second Higgs doublet $\chi$ in eq.(\ref{add})
originates from ${\overline {16}_H}(-)$, carries odd matter parity, and 
aquires the 
status of an inert doublet \cite{des, barb}.
 In addition to the Yukawa interaction in eq.(\ref{t3yuk}), 
 the following part of $SO(10)$-invariant Higgs potential is
 responsible for the radiative seesaw
\ba
V^U_{Higgs}&=&m_{10}^2
{10}_H^2+m_{16}^2{\overline{16}_H}{16}_H+\lambda_{10}{10}_H^4
+\lambda_{16}({\overline{16}_H}{16}_H)^2+
\lambda_{\rm m}{\overline{16}_H}{16}_H{10}_H{10}_H
\nonumber\\
&&+ ({\lambda_{\rm g}}/{{\rm M}_{\rm Pl}}){\overline{16}_H}{10}_H.{\overline{16}_H}
    {10}_H.{\overline {126}_H}.
\label{hspot}
\ea
This leads to the low-scale  Higgs potential 
\ba
V&=&m_{\phi}^2\phi^{\dagger}\phi+m_{\chi}^2{\chi}^{\dagger}{\chi}+\frac{1}{2}\lambda_{\phi}(\phi^{\dagger}\phi)^2+\frac{1}{2}\lambda_{\chi}({\chi}^{\dagger}{\chi})^2+\lambda_1 (\phi^{\dagger}\phi)({\chi}^{\dagger}{\chi})
+\lambda_2(\phi^{\dagger}\chi)(\chi^{\dagger}\phi)\nonumber\\
&&+\frac{1}{2}\lambda_3[(\phi^{\dagger}\chi)^2+H.c.],~~~~~\lambda_3 =
\lambda_{\rm g}<\Delta_R>/{\rm M}_{\rm Pl}. 
\label{lspot}
\ea
where ${\rm M}_{\rm Pl}= 1.2\times 10^{19}$ GeV. In the presence of
precision unification with the SM gauge symmetry  below the GUT scale, allowed 
natural value of $<\Delta_R>\sim {10}^{16}$ GeV. With 
$\lambda_{\rm g} \sim O(1)$, the embedding of the radiative seesaw mechanism
in this $SO(10)$ model then leads to the desired value of the quartic
coupling, $\lambda_3 \simeq  10^{-5}-10^{-3}$, covering the assumed value in 
~ref. \cite{rad3}. The expression for $\lambda_3  \propto <\Delta_R>$
in eq. (\ref{lspot}) also serves as an anchor to type-I and type-II seesaw
formulas. Thus, with the replacement of the externally imposed discrete
symmetry of ref. \cite{rad3} by the intrinsic matter parity (${\rm P_M}$) 
and with the replacement of RH neutrinos  of  ref.\cite{rad3}
by adjoint fermions of this model,
 $(N_1, N_2, N_3)
\to ( F_{\sigma}, F_b^1, F_b^2)$, the radiatve seesaw mechanism emerges naturally. 

Denoting ${\rm M}_{\chi_R}~({\rm M}_{\chi_I})$ as the mass of the real
 (imaginary) part of
 $\chi^0$,~it turns out that 
${\rm M}_{\chi_R}^2-{\rm M}_{\chi_I}^2
= 2\lambda_3v^2$ while the charged component mass is 
 ${\rm M}_{\chi^{\pm}}^2={\rm M}_{\chi}^2+\lambda_1v^2$.
Under the assumption that 
   ${\rm M}_{\chi_R}^2-{\rm M}_{\chi_I}^2 \ll {\rm M}_0^2=
({\rm M}_{\chi_R}^2+{\rm M}_{\chi_I}^2)/2$, which is easily satisfied 
because of the model prediction on the smallness of $\lambda_3$, the loop mediated radiative
contribution is the same as in the derivation of Ma
\cite{rad3}
\ba
({\rm M}_{\nu}^{\rm rad})_{\alpha\beta}&=& \frac{\lambda_3v^2}{8\pi^2}
{\bf \Sigma}_i\frac{y_{\alpha i}y_{\beta i}F({\rm M}_i^2/{\rm M}_0^2)}{{\rm M}_i^2},
\label{radseesaw}
\ea 
where $F(x)= [\lambda_3 v^2/(8\pi^2)][x/(1-x)]\left[1+x\ln x/(1-x)\right]$. The
formula in eq.(\ref{radseesaw}) has been noted to give the resulting seesaw formulas in three limiting cases
\ba
{{\rm M}}_{\nu}^{\rm rad}&=&\frac{\lambda_3}{8\pi^2}\left[m_a\frac{1}{{\rm M}}m_a^T, 
~~\frac{m_a}{{\rm M}_0}{\rm M}(\frac{m_a}{{\rm M}_0})^T,~~m_a\frac{1}{\Lambda}m_a^T \right],\label{lim}
\ea 
where the first, second, and the third entries  hold for
${\rm M}_i^2 \simeq {\rm M}_0^2, ~{\rm M}_i^2 \ll {\rm M}_0^2$, and $~{\rm
  M}_i^2 \gg {\rm M}_0^2$, respectively,
and we have 
defined  $m_a=yv$, ${\rm M}= {\rm diag}({\rm M}_1, {\rm M}_2, {\rm M}_3)$ , 
 $\Lambda_j={\rm M}_j[\ln({\rm M}_j^2/{\rm M}_0^2)-1]^{-1}$, and  
$\Lambda= {\rm diag}(\Lambda_1, \Lambda_2, \Lambda_3)$.
In general the neutrino mass matrix has a richer structure in this model
due to tree-level and radiative seesaw contributions
\be
{\rm M}_{\nu} = {\rm M}_{\nu}^{I}+ {\rm M}_{\nu}^{II}+{\rm M}_{\nu}^{\rm rad},\label{numass}
\ee
where the three terms on the RHS are given by eq.(\ref{seesaw12}) and  
eq.(\ref{lim}).

\noindent{\large\bf Comparison and dominance.} 
The most natural value of ${\overline {126}_H}$-Yukawa coupling to fermions is
expected to be $f\simeq 1$  which imparts substantial contribution also to
charged fermion masses near the GUT-scale \cite{baburnm, dutta} derived from their low
enegy values by renormalization group evolution \cite{dp}. 
 In the present precision unification 
model, using  ${\rm V}_{R} \simeq {\rm M}_{{\Delta}_L} \simeq {\rm M}_U \simeq
10^{16}$ GeV in  eq.(\ref{seesaw12}), we have for the third light neutrino
mass, $m_{3}^{I} \ll m_{3}^{II} \sim 10^{-3}\lambda$ eV, which is at least
one order smaller than the experimental value. Although the type-I 
contribution 
 with fine-tuned value, $f \sim 0.01$, can yield the right order of
 neutino masses, its contribution to charged fermion masses is substantially
weakened. There are SUSY $SO(10)$ models  where type-II seesaw dominance has
been shown to fit the charged fermion and neutrino sectors reasonably well 
\cite{dutta}. In the present non-SUSY model the radiative seesaw can completely
dominate and fit the available neutrino oscillation data. Compared to conventional $SO(10)$ models, the tension on $f$ and other Yukawa couplings caused due to fitting the Dirac-neutrino masses and large
neutrino mixings  is absent in the present model. 
As a result, the
Yukawa couplings of Higgs representations ${\overline {126}_H}$,
${10}_H$, and ${120}_H$ get almost decoupled from the neutrino sector 
 with their full potential to parametrize the charged fermion masses and mixings in a much more effective
 manner. This is  natural as the radiative seesaw is
basically designed to be more dominant as it admits much lighter seesaw scale.
While details of these and 
a number of new $SO(10)$ applications \cite{rad5} will be reported
elsewhere, we confine here to the triplet fermion DM discussed earlier using SM
extensions \cite{cirelli,rad4}. 
 
\par\noindent Depending upon their actual masses, the Yukawa interaction in
 eq.(\ref{t3yuk}) introduces decays, $\chi \to F_{\sigma} F^i_b$,
$F_{\sigma}\to l{\bar l}F^i_b$, or  $F^i_b \to l{\bar l}F_{\sigma}$ for
 $i=1,2$. Then only
 the  lightest of them can be a stable dark matter candidate. Thus
the model offers the possibility of fermionic weak triplet, singlet, or inert
scalar doublet as a  CDM candidate.    
In eq.(\ref{seesaw12}) the Dirac neutrino mass, being of the same order as
 the up quark mass, the experimental value of large 
top-quark mass pushes the canonical seesaw scale closer to the 
GUT scale. As there is no such constraint on the Yukawa couplings in
 eq.(\ref{t3yuk}) and eq.(\ref{lim}), especially from experimental data, they
 can be small
as has been assumed in \cite{rad3}. Since the model permits additional lepton
flavor violating processes compared to conventional SO(10) models, these
 couplings are constrained by $\mu \to e\gamma$  and other decay rates.  
For example, in order to have the triplet fermionic DM,
 the two adjoint singlet fermions and
 the inert scalar doublet are needed to be heavier than the triplet fermion 
  and we examine this possibility 
 including the new T2K  data \cite{he}. Denoting  ${\rm M}_1=m_{F_{\sigma}}, 
{\rm  M}_2= m_{F_b^1}$, and ${\rm M}_3=m_{F_b^2}$, 
we choose ${\rm M}_1 < {\rm M}_2
<  {\rm M}_3$ with ${\rm M_i}^2 \ll m_{\chi}^2$ for which the  
 second relation of eq.(\ref{lim}) applies. 
 Using neutrino mixing angles $\theta_{12}= 33^o, \theta_{23}= 43^o$, 
 $\theta_{13}=10^o$, and all phases to be
vanishing, we have the mixing
matrix elements ${\rm U}_{ei}=(0.808,0.555,0.190), {\rm U}_{\mu i}=(-0.661,0.530 ,0.666)$, and
${\rm U}_{\tau i}=(0.269,-0.638,0.719)$. Using the relation $y_{\alpha i} = 
{\rm U}_{\alpha i}y_i$ , the
$\mu \to e\gamma$ decay rate constraint becomes $|(|y_1|^2 -(2/3)|y_2|^2 -(2/7)
|y_3|^2)| \le
0.672 ~(m_{\chi}/ 2.7 ~{\rm TeV})^2$ which is different from the 
 tribimaximal mixing constraint \cite{rad4}. With $\lambda_3 \sim
 O(10^{-5})$ and $y_i \sim O(10^{-2})$, the
 desired neutrino  mass eigen
values , $m_i = (0.010,0. 0135, 0.050 )$ eV,
in the hierarchial case are obtained 
for ${\rm M}_i= (2.7, 3.0, 3.3)$ TeV and $m_{\chi}= 40$ TeV . But we note that 
while all other parameters and predicted masses remain unchanged, the mass of
 the inert doublet is brought down to $m_{\chi} \sim 5$ TeV  for $y_i \sim
 O(10^{-3})$. Inspite of nearly one order reduction in  $m_{\chi}$,
 the muon decay constraint is very well satisfied because of corresponding 
smaller values of
 the allowed Yukawa couplings. It is interesting to note that these couplings, $y_i\sim 10^{-2}(10^{-3})$,
with adjoint fermions are of the same order as the charged-lepton Yukawa 
couplings for $\tau^{-}(\mu^{-})$ in the non-SUSY SM.    
We also obtain solutions consistent with inverted hierarchy for  
$y_1 \le y_2  \le y_3$ in each case. 
   The small values of $y_i$ used here do not cause any problem since more rapid rate of annihilation and
 coannihilation required to produce right value of relic density of the triplet
fermionic DM is
 accomplished by gauge boson interactions \cite{cirelli,rad4}. We find that 
 the masses of all the particles, mediating the radiative seesaw or
 responsible for its low scale,   
are in the range accessible to LHC or planned colliders.

\noindent{\large\bf Experimental signatures.}
It would be worthwhile to discuss some of the possible experimental signatures of this
model which may distinguish it from the other non-SUSY GUT-based radiative
seesaw model with $SU(5)\times Z_2$ grand unification symmetry \cite{rad4}.

\noindent{\large\bf (a)Proton lifetime predictions: $SO(10)$ vs. $SU(5)$. } 
In order to make a possibly clear distinction we discuss
 gauge boson
mediated proton decay  $p \to e^+\pi^0$
for which there are ongoing dedicated experimental searches \cite{nishino}
with measured value of the lower limit on the life time, 
$\tau_p^{expt.}~\ge ~1.01\times 10^{34}$~~yrs.        
With a choice of TeV scale particle spectrum different from
eq.(\ref{add}), unification of couplings has been obtained in ref.\cite{rad4}
with  
$M_U^{SU(5)}=2.65\times 10^{15}$ GeV and the  approximation adopted appears to
predict proton lifetime substantially lower than the current experimental
limit. Noting that the model prediction for the actual inverse decay rate has been
underestimated, we re-evaluate it while estimating proton lifetime prediction in the
present $SO(10)$ model. Including strong and electrowek renormalization
effects on the ${\rm d}=6$ operator and taking into account quark mixing, chiral symmetry breaking
effects, and lattice gauge theory estimations, the decay rates
 for the two models are \cite{bajc, babupati}, 
\ba   
\Gamma(p\rightarrow e^+\pi^0) 
&=&\frac{m_p}{64\pi f_{\pi}^2}
\frac{{g_G}^4}{{M_U}^4})|A_L|^2|\bar{\alpha_H}|^2(1+D+F)^2\times R,\nonumber\\
\label{width}
\ea
where $ R=[A_{SR}^2+A_{SL}^2 (1+ |{V_{ud}}|^2)^2]$ for $SU(5)$, but $R=
[(A_{SR}^2+A_{SL}^2) (1+ |{V_{ud}}|^2)^2]$ for $SO(10)$, $V_{ud}=0.974=$ 
 the  $(1,1)$ element of $V_{CKM}$ for quark mixings, and
$A_{SL}(A_{SR})$ is the short-distance renormalization factor in the
left (right) sectors.  In eq.(\ref{width}) $A_L=1.25=$
long distance renormalization factor which is the same for both models, but 
$A_{SL}\simeq A_{SR}=2.414 ~(2.542)$ for $SU(5) ~(SO(10))$,  
 $M_U=$ degenerate mass of $12 ~(24)$ superheavy gauge bosons in
$SU(5)~(SO(10))$, $\bar\alpha_H =$ hadronic matrix elements, $m_p =$proton mass
$=938.3$ MeV, $f_{\pi}=$ pion decay 
constant $=139$ MeV, and the chiral Lagrangian parameters are $D=0.81$ and
$F=0.47$. With $\alpha_H= \bar{\alpha_H}(1+D+F)=0.012$ GeV$^3$ estimated from 
lattice
gauge theory computations, we obtain  $A_R \simeq A_LA_{SL}\simeq
A_LA_{SR}\simeq 3.02 ~(3.18)$ for $SU(5) ~(SO(10))$, and the expression for the
 inverse
decay rates for both the models is,   

\ba
\Gamma^{-1}(p\rightarrow e^+\pi^0)
& = & \frac{4}{\pi}\frac{f_{\pi}^2}{m_p}\frac{M_U^4}{\alpha_G^2}\frac{1}{\alpha_H^2 A_R^2}\frac{1}{F_q},\label{taup}
\ea
where the GUT-fine structure constant $\alpha_G=1/38.25$ and the
factor $F_q=1+(1+|V_{ud}|^2)^2\simeq 4.8$ for $SU(5)$, but  $\alpha_G=1/35.3$
and $F_q=2(1+|V_{ud}|^2)^2\simeq 7.6$ for $SO(10)$. This formula  reduces to
the form given in \cite{psb,babupati} and sets the lower limit for non-SUSY
SU(5) GUT scale to be $M_U\ge 10^{15.5}$ GeV from the experimental lower limit
on $\tau_p$.   
 Now using  the estimated values of the model parameters in each
 case eq.(\ref{taup}) gives,
\ba
\tau_p^{SU(5)XZ_2}&\simeq& 6.26\times 10^{33}~~yrs,\nonumber\\
\tau_p^{SO(10)}&\simeq&4.28\times 10^{35}~~yrs.\label{taupnum}
\ea    
which mark clear distinction between the two models with much greater proton
stability in the present model due to  larger unification scale that
originates from its TeV-scale spectrum.
Thus, we have improved the proton lifetime 
estimation of \cite{rad4} in the $SU(5)\times Z_2$ model by at least one
order. The significance of this estimation is that the small deficit from the
experimental lower limit can be compensated 
by invoking small threshold effects at the GUT-scale or the TeV scale.
But there is a possibility that this model would be constrained if the
future measurements increase the existing lower limit by about one order or
 longer.

In the present model, however, the one-loop prediction of proton lifetime is nearly
$40$ times longer than the current limit which may be accessed by 
the search experiments of the next generation\footnote{As discussed in \cite{psb}
the proton lifetime prediction in this model can also be reduced to nearly half of its predicted value  when the scalar mediators in ${\overline{10}_H}\subset SU(5)$
  contained in ${\overline{16}_H}\subset SO(10)$ have masses near $\sim$ TeV
  scale needed
  for shorter lifetime of cosmologically safe color-octet
  fermion \cite{baer}.}. 
The two-loop and small 
  threshold
effects \cite{parpat} may also bring this prediction closer to the current experimental limit
without disturbing precision unification.
Since the TeV scale spectrum in the present model is
richer and also two of the mediating fermions for
radiative seesaw are different from the two RH neutrinos of ref.\cite{rad4}, there would
be a variety of other possible ways by which this  model can be experimentally
distinguished from others. Pending these and other related
investigations \cite{mkprep}, the present estimation shows that improvement on the
existing proton
lifetime measurement by at least upto one order \cite{nishino}
would clearly favor the present $SO(10)$ based radiative seesaw model of
precision unification.
 
\noindent{\large\bf(b)Accelerator tests.} It has been shown that the
color-octet fermion present in the TeV-
scale spectrum would be pair produced  at the LHC with nearly 1000, 15, and 
 2.5 number of events for its mass $m_{F_C}=2.0$ TeV, $3.0$ TeV, and $3.5$
TeV, respectively, with  $100$fb$^{-1}$ beam luminousity  at energy $\sqrt s=14$
TeV \cite{psb}. ~Its relic density problem can be evaded
either non-perturbatively, or by invoking second inflation \cite{baer}, or
even
perturbatively, by making its lifetime shorter \cite{arkani} through  the
introduction of the 
complete Higgs multiplet ${10}_H \subset SU(5)$ contained in ${16}_H\subset
 SO(10)$ at an appropriate mass scale without disturbing precision 
unification. 
In the third case, there is a clear possibility of the boosted production of
CDM candidates $F_{\sigma}$ or $F_b$ with displaced vertices at
LHC  via $F_C{\bar {F}}_C$ pair production \cite{mkprep}.   
Alternatively, the octet fermion
can be replaced by a pair of  color octet scalars to achieve the same
precision unification and their LHC signatures have been discussed in detail
\cite{dobrescu}. They can be pair produced copiously at LHC energies and will
manifest themselves as resonances in multijet final states. Another specific
distinguishing signal at LHC or Tevatron, but more prominent at ILC, would be  the pair production and decay of heavy  charged fermions contained
in $F_{\phi}$ and $F_{\chi}$ of the non-standard spectrum of the extended 
SO(10) model \cite{mkprep}.

\noindent{\large\bf Summary and conclusion.}
While discrete symmetries are externally imposed on
model extensions to maintain stability of incorporated dark matter,
a minimal non-SUSY $SO(10)$ model naturally possess the 
stabilizing matter parity discrete
symmetry, but it does not unify gauge couplings and neither does it predict
prospective DM candidates. Although 
 it predicts very attractive neutrino mass generation mechanisms, they
 involve  high
seesaw scales, $10^{13}-10^{15}$ GeV, inaccessible for experimental tests 
in foreseeable future.  
Here the model is
successfully extended to realize a
low scale spectrum that achieves precision unification with
experimentally testable
proton lifetime, predicts an inert scalar
doublet, and other potential fermionic DM candidates. With matter parity
conservation, the type-I and type-II seesaw formulas are automatic consequences 
of the model, but
mediated by the DM
and the inert doublet, most interestingly,
it also predicts the verifiable low-scale radiative seesaw
formula of Ma type along with the natural emergence of its small
quartic coupling. Moreover, the new
contribution to neutrino mass dominates over the conventional ones 
making a major impact  on
neutrino physics and dark matter while 
opening up high potential as a theory of fermion masses in general. Our
estimation  in the $SU(5)\times Z_2$  based radiative seesaw model reveals the proton
lifetime for $p\to e^+\pi^0$ to be somewhat less than 
the current  experimental lower 
limit, but in the present $SO(10)$ model the lifetime turns out to be
nearly $40$ times longer which marks one of its clear distinguishing features. 

\noindent{\large\bf Acknowledgment.}
M.K.P. thanks Harish-Chandra Research Institute for a visiting position.

\end{document}